\documentclass[11pt,twoside]{article}
\usepackage{newpasp}

\usepackage{epsf}

\index{Mihos, J.C.}

\begin{document}

\title{Gas/Star Offsets in Tidal Tails}
\author{J. C. Mihos}
\affil{Case Western Reserve University, Cleveland, OH 44106}


\begin{abstract}
We use numerical simulations to study the development of gas/star
offsets in the tidal tails of merging galaxies. These offsets are
shown to be a natural consequence of the radially extended HI spatial
distribution in disk galaxies, coupled with internal dissipation in
the gaseous component driven by the interaction. This mechanism
explains the observed gas/star offsets in interacting galaxies without
invoking interactions with a hot (unseen) gaseous component.
\end{abstract}




\section{Observational Motivation}

The development of tidal tails during galaxy encounters is
predominantly a gravitational effect; as a result, both gas
and stars
should have similar kinematics and spatial morphology on large
scales. However, in a significant number of systems the HI and stellar
tidal debris do in fact show marked spatial offsets, including the
tidal tails of NGC 520 (Hibbard, Vacca, \& Yun 2000), 7714/5 (Smith
etal 1997), and, most dramatically, NGC 3690 (Hibbard etal 2000). At
face value, these observations suggest that
hydrodynamic forces may also play a role in the global evolution of
the tidal tails, via effects such as an interaction between the tidal
gas and a hot gaseous halo or starburst wind (Hibbard etal 2000). Here
I show that the offsets arise naturally from the differing
radial distributions of gas and stars, coupled with interaction-driven
dissipation and inflow in the tidal gas. No hot gas component is needed
to explain these offsets.


\begin{figure}
\plotone{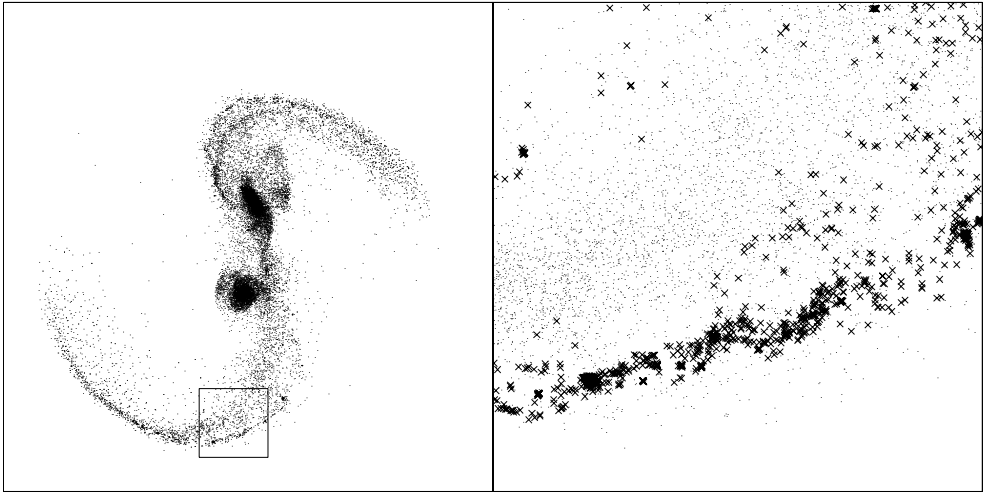}
\caption{\small The left panel shows the simulation viewed just prior
to merging. The box in the bottom tail shows the region which is
magnified at right. In the magnified panel, the stars a shown as dots
and the gas particles as crosses. A substantial offset between the two
components is seen.}
\label{fig1}
\end{figure}

 

\section{Dynamical Modeling}

Previous dynamical modeling of merging galaxies (e.g., Barnes \&
Hernquist 1991, 1996; Mihos \& Hernquist 1996) showed no evidence for
segregation of gas and stars in tidal debris. However, in those
models, both the gas and stars had identical exponential spatial
distributions, whereas the gas distribution in disk galaxies is
typically much more spatially extended than the starlight. I have
rerun the fiducial merger model of MH96, using an gas distribution
which is exponential within $2.5h$ (where $h$ is the stellar disk
scale length), then falls as $r^{-1}$ until a truncation radius
of $r_t=8h$. Figure 1 shows the model just prior to the final merging,
and a prominent gas/star offset can be seen in the prograde tail.
The fact that we see an offset at all argues that these offsets
need not be the result of an interaction between the tail gas 
and a hot gas component, since no such hot gas component is present
in our model.

Is the offset due solely to the differing initial spatial distribution
of the gas and stars? A simulation was run identical to the first,
but with the gas particles evolved collisionlessly. In this
case, no gas/star offset developed, indicating that hydrodynamic
forces are at work in shaping the tails. By analyzing a subset of tail
particles in the full hydrodynamical model, we find that both the gas
and star particles seen in the tail are following a largely
collisionless evolution on large scales. The gas in the tail comes
from larger initial radius than the stars; its systematically larger
angular momentum, results in a leading gas tail. So what happened to
the gas that would have been cospatial with the stars?

To answer this question, we followed the evolution of with gas
particles that came from the same region of initial (i.e.,
pre-encounter) phase space as the stars which ended up in the tidal
tail. Because of the exponential distribution of stars, these gaseous
``phase space partners'' come from small radius in the disk and are
more subject to collisional dissipation and inflow during the
encounter. As a result, this gas decouples from the collisionless
evolution that the stars follow, flowing inwards and leaving the
stellar tail gas-poor.  In essence, it is not the hydrodynamic
evolution of the gas {\it observed} in the tail which leads to the
offset, but rather the hydrodynamic decoupling and inflow of the low
angular momentum gas which would have otherwise filled in the stellar
tail.

\end{document}